\documentclass[aps,pra,twocolumn,10pt,superscriptaddress,showpacs,amsmath,amssymb,nofootinbib]{revtex4-1}
\usepackage{graphicx}
\usepackage{hyperref}
\usepackage{amssymb}
\usepackage[margin=1in]{geometry}

\begin{document}

\title{Learning about non-Newtonian fluids in a student-driven classroom}
\author{D.~R.~Dounas-Frazer}
\email{drdf@berkeley.edu}
\affiliation{Department of Physics, University of California at
Berkeley, Berkeley, California 94720-7300, USA}
\author{J.~Lynn}
\affiliation{Department of Physics, University of California at
Berkeley, Berkeley, California 94720-7300, USA}
\affiliation{Theoretical Astrophysics Center, University of California at
Berkeley, Berkeley, California 94720-7300, USA}
\author{A.~M.~Zaniewski}
\affiliation{Department of Physics, University of California at
Berkeley, Berkeley, California 94720-7300, USA}
\affiliation{Materials Science Division, Lawrence Berkeley National Lab, Berkeley, California 94720, USA}
\affiliation{Center of Integrated Nanomechanical Systems, University of California at
Berkeley, Berkeley, California 94720-7300, USA}
\author{N.~Roth}
\affiliation{Department of Physics, University of California at
Berkeley, Berkeley, California 94720-7300, USA}
\affiliation{Theoretical Astrophysics Center, University of California at
Berkeley, Berkeley, California 94720-7300, USA}
\date{\today}

\maketitle

We describe a simple, low-cost experiment and corresponding pedagogical strategies for studying fluids whose viscosities depend on shear rate, referred to as \emph{non-Newtonian fluids}. We developed these materials teaching for the Compass Project~\cite{Compass}, an organization that fosters a creative, diverse, and collaborative community of science students at UC Berkeley. Incoming freshmen worked together in a week-long, residential program to explore physical phenomena through a combination of conceptual model-building and hands-on experimentation. During the program, students were exposed to three major aspects of scientific discovery: developing a model, testing the model, and investigating deviations from the model.

We chose to study non-Newtonian fluids because they are not only the subject of active research~\cite{Cheng2011}, but they are also ubiquitous in our everyday lives; examples include ketchup, toothpaste, and sand. Despite its importance, fluid dynamics is a subject that is typically neglected in high school physics curricula. Nevertheless, non-Newtonian fluids are amenable to a wide range of engaging, collaborative activities~\cite{Habdas2006}.

One of our goals was for students to derive \emph{Stokes' Law}, which describes the velocity dependence of the drag force experienced by a ball moving in a viscous fluid. Stokes' Law, which is valid when the flow is sufficiently viscous that no turbulence develops, is~\cite{Bat67}:
\begin{equation}
  \label{eq:stokes}
  F = 6 \pi C R \eta v,
\end{equation}
where $R$ and $v$ are the ball's radius and speed, respectively, $\eta$ is the fluid's viscosity, and $C$ is a dimensionless constant that depends on the geometry of the fluid's container ($C=1$ when the container is very large compared to the size of the ball)~\cite{Ambari1985}. Fluids for which $\eta$ is independent of $v$ are known as Newtonian, whereas those for which it is not are non-Newtonian. Thus, for Newtonian fluids, Stokes' Law predicts a linear relationship between the speed of the ball and the drag force it experiences. Non-Newtonian fluids, on the other hand, deviate from Stokes' Law, often by exhibiting a nonlinear dependence of $v$ on $F$.  We emphasize that this distinction is only valid when the Reynolds number is low and the flow is laminar.  At high Reynolds numbers, turbulence spoils the linear relationship between $v$ and $F$ for Newtonian fluids~\cite{Bat67}.

We discuss how our students derived (\ref{eq:stokes}), and describe an experiment that they used to measure $v(F)$ for two different fluids: corn syrup (Newtonian) and a mixture of cornstarch and water called \emph{oobleck} (non-Newtonian). The apparatus we present is different from others~\cite{Seckin1996,Dolz2005,Courbin2005,Digilov2008} that discriminate between Newtonian and non-Newtonian fluids in that it is relatively simple and low cost. In contrast to falling-ball viscometers~\cite{Brizard2005}, students do not need to consider buoyant forces to understand the operation of our apparatus.  Finally, students examined oobleck under a microscope, thereby illustrating the origins of oobleck's non-Newtonian behavior. Over the course of the week, students spent a total of 10 hours in the classroom engaging in small- and large- group discussions, 3 hours performing experiments, and about 6 hours completing collaborative homework assignments; we present a somewhat streamlined account of this work.

\section{Deriving Stokes' Law}

To guide the students' derivation of Stokes' Law, we followed a classroom model called Modeling Discourse Management (MDM)~\cite{Desbien2002}, and incorporated the method Think-Pair-Share \cite{Lyman1981}.  Students worked in groups of four, using a small whiteboard to record their work and share ideas with the class. MDM prescribes that if the class is missing a critical or interesting idea in their small group discussions, teachers should attempt to ``seed''  discussions by asking simpler, related questions that could serve as stepping stones to the main result.

We introduced the concept of viscosity by asking students to describe what happens to a thick fluid between two parallel plates separated by distance $\Delta y$ when one plate moves at constant speed $\Delta v$ relative to the other, fixed plate. An example of a question we used to seed discussion in one small group was, ``What is the fluid right next to the plate doing?'' Students correctly intuited a flow pattern with a position-dependent velocity that increases smoothly from zero at the stationary plate to $\Delta v$ at the moving plate, a pattern commonly referred to as \emph{Couette flow}~\cite{Bat67}, and that there must be forces exerted both between neighboring layers of fluid and at the plate-fluid interface. Additionally, students guessed that processes at the molecular level must be responsible for this ``sticking'' or ``dragging''.  We took the standard approach of lumping all of the microscopic details into one macroscopic quantity, the viscosity $\eta$.

We asked students to identify the parameters on which the force exerted on the top plate depends, and to articulate how the force depends on those factors. Students identified four relevant dependencies, namely: $F$ increases with increasing $\eta$, $\Delta v$, and plate area $A$, and $F$ decreases with increasing $\Delta y$.  We then asked students to construct a quantitative expression for the drag force on one fluid layer due to a neighboring layer using dimensional analysis, a technique that has been used to study other aspects of fluid flow~\cite{Guerra2011}. Given that viscosity has units of Pa$\cdot$s, students successfully constructed the only simple expression that satisfies the above dependencies:
\begin{equation}
  \label{eq:simpleDrag}
  F = \eta A \Delta v/ \Delta y,
\end{equation}
which reduces to Newton's law of viscosity in the limit $\Delta v/\Delta y\rightarrow dv/dy$~\cite{Bat67}.

Next, we asked the students to use (\ref{eq:simpleDrag}) to make a prediction about the drag force on a ball of radius $R$ moving at velocity $v$ through an otherwise stationary fluid that is infinite in extent. Students made the following approximations (assisted by seed questions as necessary): the area over which the force is applied is approximately the surface area of the ball, so $A\approx 4\pi R^2$; because the fluid far from the ball is stationary whereas the fluid near the ball moves with speed $v$, the change in velocity is $\Delta v = v$; and, since $R$ is the only available length scale, the characteristic distance over which the change in velocity occurs is $\Delta y \approx R$. These approximations yield
\begin{equation}
  \label{simpleStokes}
  F \approx 4 \pi R \eta v,
\end{equation}
consistent with Stokes' Law. Because determination of $C$ in (\ref{eq:stokes}) was beyond the scope of our goals, we focused on the scaling of $F$ with $v$ rather than the absolute magnitude of the force.

\begin{figure}\center \includegraphics[width=\columnwidth]{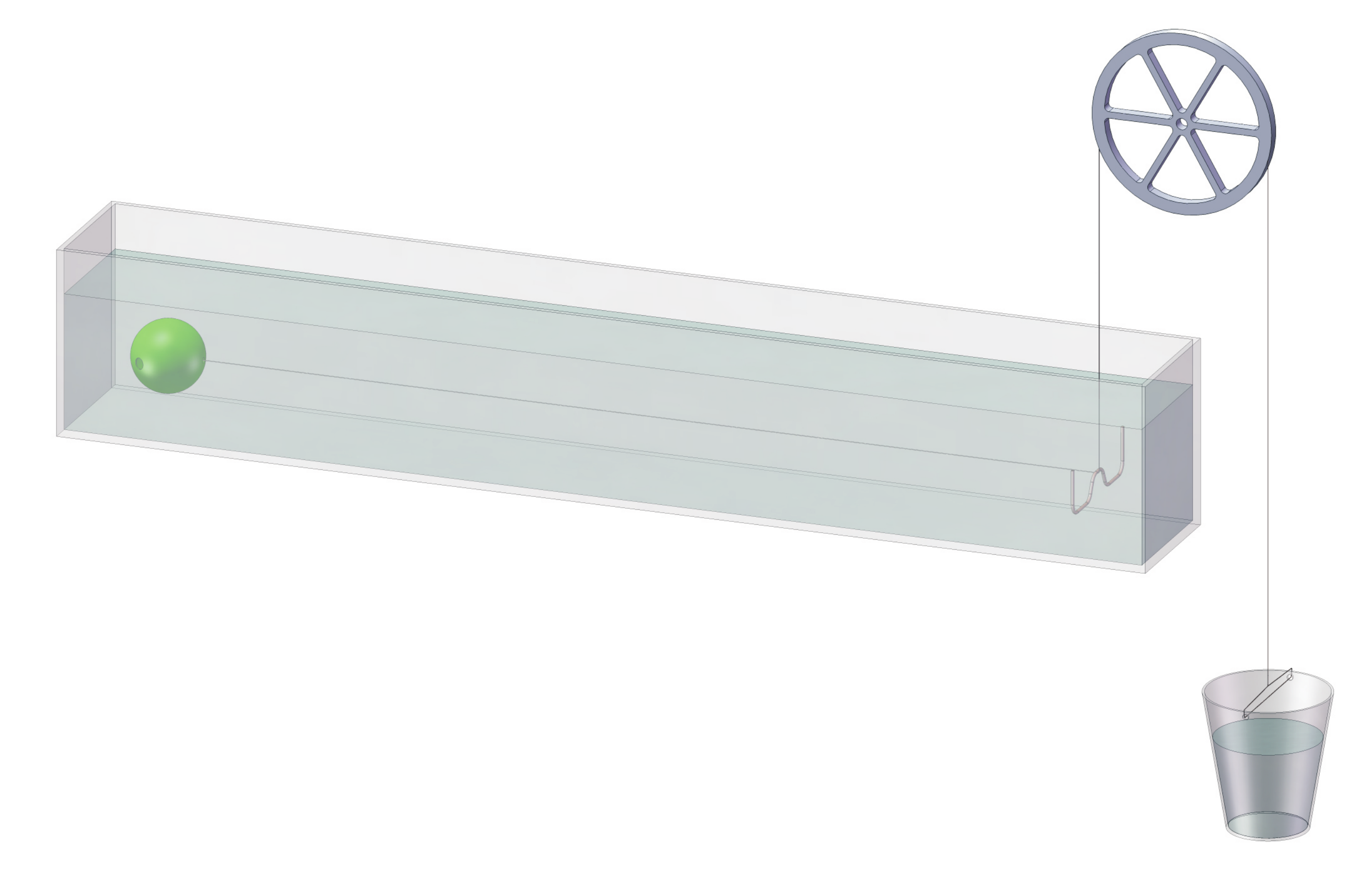}
\caption{\label{fig:apparatus} Schematic of appartus used to measure velocity dependence of fluid drag forces.}
\end{figure}

\begin{figure}\center \includegraphics[width=\columnwidth]{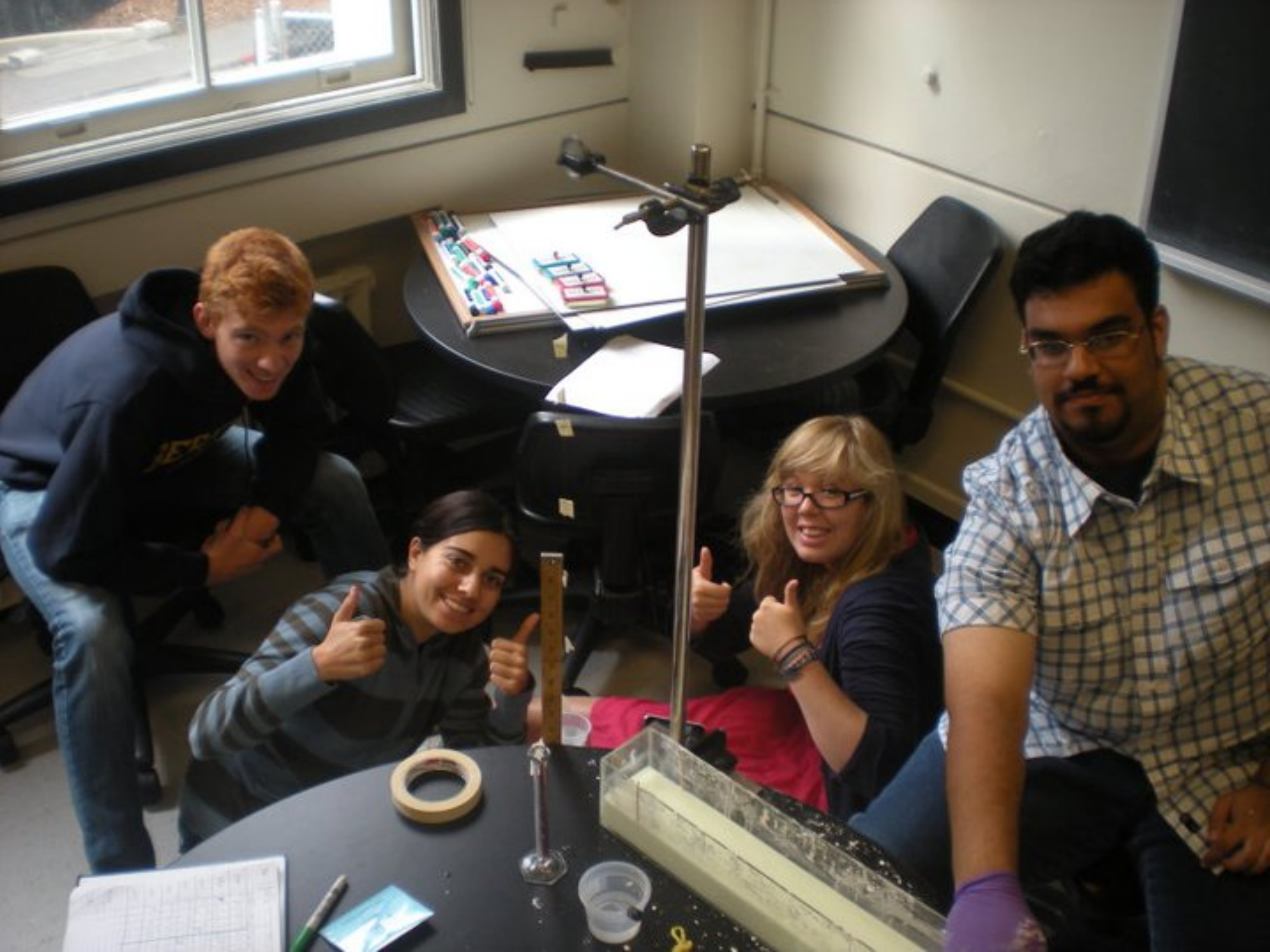}
\caption{\label{fig:students} Student experimentalists measuring drag forces in oobleck.}
\end{figure}

\section{Apparatus and procedure}

The apparatus, shown in Fig.~\ref{fig:apparatus}, consists of a ping-pong ball submerged in a fluid-filled, acrylic trough (60~cm long, 8~cm wide, and 10~cm tall), with a wire that connects the ball to a counterweight of mass $m$ via a crossbeam and pulley.  When the weight is released from rest, the ball accelerates to a constant (terminal) velocity $v$, at which point the counterweight exactly balances the drag force experienced by the ball: $F=mg$, where $g=9.8$~m/s$^2$ is the standard gravity on Earth. Because the ball and counterweight are attached by a taut wire, they travel at the same speed. Thus $F$ and $v$ can be determined by measuring the mass and speed of the counterweight. The counterweight was a 3~g cup filled with up to 60~mL water, which allowed for an easily scalable system.

Three holes were drilled in the ball, a small hole of diameter 3~mm and two large ones of 6~mm. The purpose of the holes was twofold: first, to facilitate attachment of the ball to the wire; and second, to allow the ball to fill with fluid in order to prevent it from floating or sinking.  The ball was attached to the wire by threading the wire through the small hole and tying it to a plastic bead 5~mm in diameter. Though only one large hole was needed to thread the bead, the second hole greatly reduced the time needed to fill the ball with the ambient fluid.

Although we, the teachers, designed and built the apparatus, students developed their own measurement procedure. The class reached consensus on the procedure, and all groups followed the same steps. To measure $v$, students marked 4 successive 10-cm intervals on the wire with flags of tape (Fig.~\ref{fig:students}), and timed them with a stopwatch as they passed the bottom edge of the trough. For a given counterweight mass, the experiment was repeated 4 times. In all their experiments, students allowed the ball to travel an initial distance of 10~cm before measuring its speed to ensure that it had reached terminal velocity. This distance was determined empirically.

\begin{figure}[t]\center \includegraphics[width=\columnwidth]{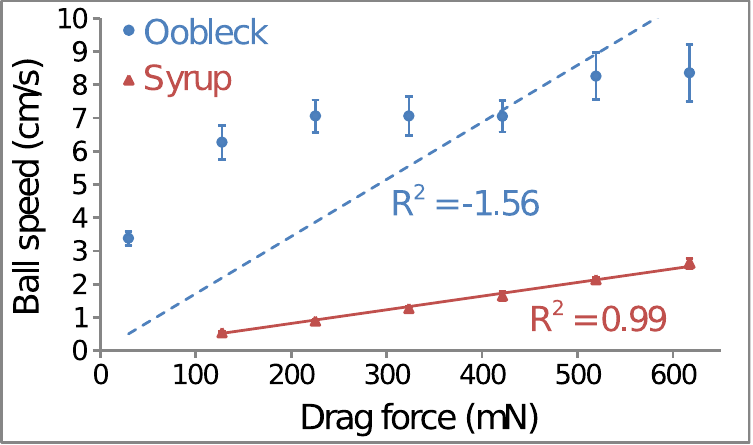}
\caption{\label{fig:data} Weighted class average of student measurements of ball speed $v$ as a function of drag force $F$ for oobleck (blue circles) and corn syrup (red triangles). The function $v(F)=\alpha F$ was fit to each data set. Fitted curves are plotted as dashed blue and solid red lines for oobleck ($\alpha = 0.2$~s/kg) and syrup ($\alpha=0.04$~s/kg), respectively. The nonlinear dependence of $v$ on $F$ in oobleck indicates that it is a non-Newtonian fluid.}
\end{figure}

\begin{figure}[t]\center \includegraphics[width=\columnwidth]{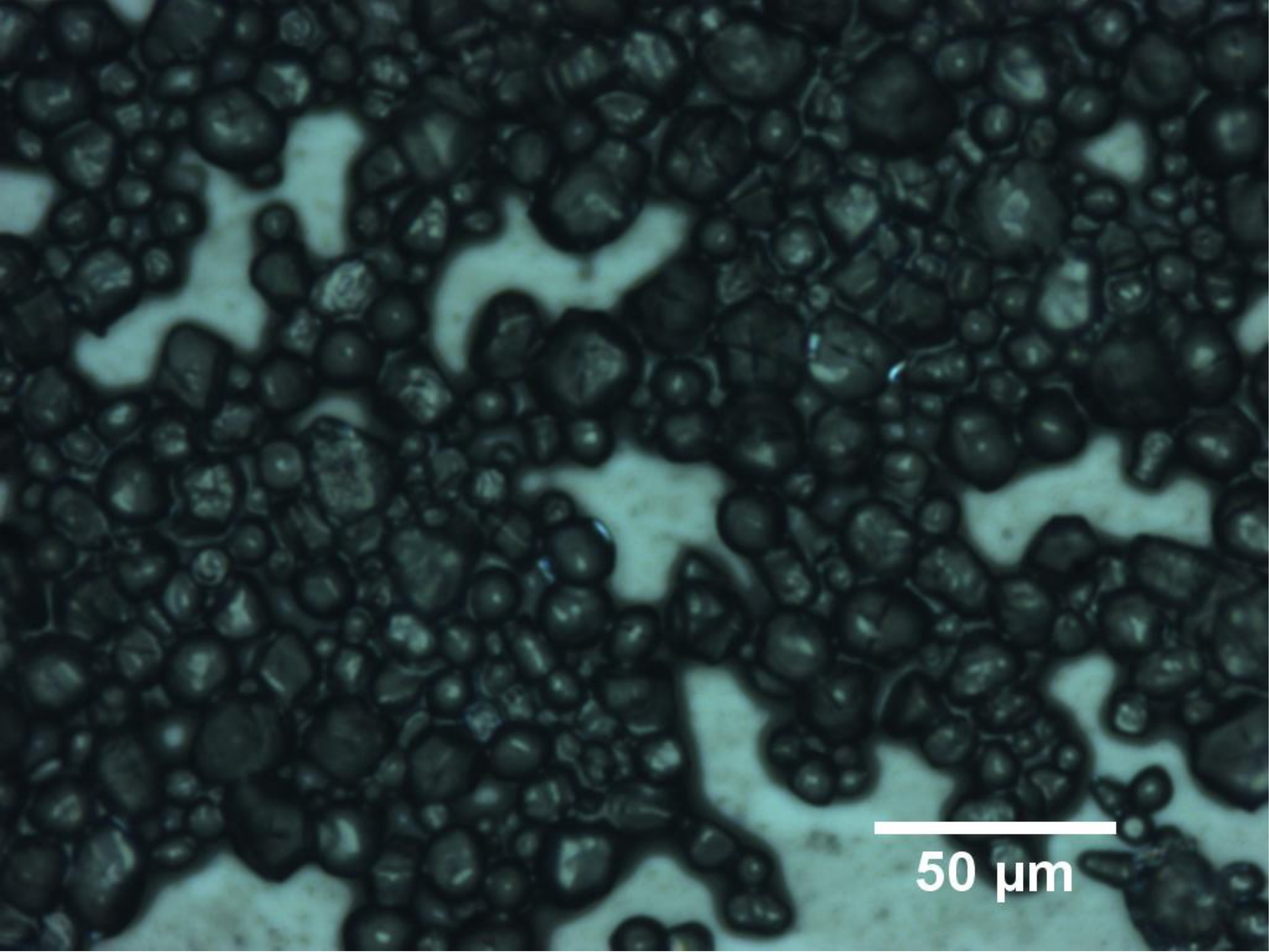}
\caption{\label{fig:microscope} Microscopic image of oobleck. The best  images were obtained by looking near the edges of a droplet or through a very thin sample.}
\end{figure}

\section{Results and discussion}

The students fit the $v(F) = \alpha F$ model to their data using Microsoft Excel, and the corresponding correlation coefficient $R^2$ was used to determine whether or not a fluid was Newtonian: $R^2 \approx 1$ for Newtonian fluids, and $R^2\not\approx 1$ indicates non-Newtonian behavior. For Newtonian fluids, the viscosity could in principle be determined from the slope $\alpha$ using Stokes' Law. To do so, the constant $C$ appearing in (\ref{eq:stokes}) must be determined for the particular apparatus~\cite{Dolz2005,Ambari1985}. However, the goal of our experiment was to test the prediction of linear relationship between $v$ and $F$, thereby discriminating between Newtonian and non-Newtonian fluids. For this purpose, it is sufficient to determine whether the data is linear. The results of the students' experiments, shown in Fig.~\ref{fig:data}, indicate that syrup is Newtonian ($R^2=0.99$) whereas oobleck ($R^2=-1.56$) is not.

In a subsequent class, we encouraged students to think about microscopic differences between oobleck and corn syrup that could result in their macroscopically different behavior.  After forming hypotheses of what they expected to see, students examined a thin sample of oobleck under a microscope. They observed that corn starch crystals were not dissolved but suspended in the water, as can be seen in Fig.~\ref{fig:microscope}.  Packing and jamming of the crystals, two mechanisms for the shear-thickening nature of suspensions~\cite{Cheng2011}, were observed by agitating the sample.

In conclusion, students developed a conceptual and formal understanding of fluid viscosity, measured viscous forces of various fluids, and explored the microscopic origins of oobleck's non-Newtonian behavior.  In addition to identifying non-Newtonian fluids, the apparatus we designed has other potential uses, \emph{e.g.}, determination of the velocity dependence of turbulent drag forces or viscometry. The pedagogical and experimental methods described here are appropriate for advanced high school and undergraduate students.

\acknowledgments The authors acknowledge helpful discussions with B.~Albanna, J.~Corbo, D.~Edelberg, A.~Little, and G.~Quan. This work was supported by the Departments of Physics, Astronomy, and Earth and Planetary Sciences at UC Berkeley, the Center for Integrative Planetary Science, and private donations. DRDF, JL, and AMZ were supported by the National Science Foundation under grants  PHY-1068875, DGE-1106400, and EEC-0832819, respectively. NR was supported by the Department of Energy Office of Science Graduate Fellowship Program, made possible by the American Recovery and Reinvestment Act of 2009, administered by ORISE-ORAU under contract DE-AC05-06OR23100.


\begin{thebibliography}{13}%
\makeatletter
\providecommand \@ifxundefined [1]{%
 \@ifx{#1\undefined}
}%
\providecommand \@ifnum [1]{%
 \ifnum #1\expandafter \@firstoftwo
 \else \expandafter \@secondoftwo
 \fi
}%
\providecommand \@ifx [1]{%
 \ifx #1\expandafter \@firstoftwo
 \else \expandafter \@secondoftwo
 \fi
}%
\providecommand \natexlab [1]{#1}%
\providecommand \enquote  [1]{``#1''}%
\providecommand \bibnamefont  [1]{#1}%
\providecommand \bibfnamefont [1]{#1}%
\providecommand \citenamefont [1]{#1}%
\providecommand \href@noop [0]{\@secondoftwo}%
\providecommand \href [0]{\begingroup \@sanitize@url \@href}%
\providecommand \@href[1]{\@@startlink{#1}\@@href}%
\providecommand \@@href[1]{\endgroup#1\@@endlink}%
\providecommand \@sanitize@url [0]{\catcode `\\12\catcode `\$12\catcode
  `\&12\catcode `\#12\catcode `\^12\catcode `\_12\catcode `\%12\relax}%
\providecommand \@@startlink[1]{}%
\providecommand \@@endlink[0]{}%
\providecommand \url  [0]{\begingroup\@sanitize@url \@url }%
\providecommand \@url [1]{\endgroup\@href {#1}{\urlprefix }}%
\providecommand \urlprefix  [0]{URL }%
\providecommand \Eprint [0]{\href }%
\providecommand \doibase [0]{http://dx.doi.org/}%
\providecommand \selectlanguage [0]{\@gobble}%
\providecommand \bibinfo  [0]{\@secondoftwo}%
\providecommand \bibfield  [0]{\@secondoftwo}%
\providecommand \translation [1]{[#1]}%
\providecommand \BibitemOpen [0]{}%
\providecommand \bibitemStop [0]{}%
\providecommand \bibitemNoStop [0]{.\EOS\space}%
\providecommand \EOS [0]{\spacefactor3000\relax}%
\providecommand \BibitemShut  [1]{\csname bibitem#1\endcsname}%
\let\auto@bib@innerbib\@empty
%</preamble>
\bibitem [{Com()}]{Compass}%
  \BibitemOpen
  \href@noop {} {}\bibinfo {note} {See
  \url{www.berkeleycompassproject.org}}\BibitemShut {NoStop}%
\bibitem [{\citenamefont {Cheng}\ \emph {et~al.}(2011)\citenamefont {Cheng},
  \citenamefont {McCoy}, \citenamefont {Israelachvili},\ and\ \citenamefont
  {Cohen}}]{Cheng2011}%
  \BibitemOpen
  \bibfield  {author} {\bibinfo {author} {\bibfnamefont {X.}~\bibnamefont
  {Cheng}}, \bibinfo {author} {\bibfnamefont {J.~H.}\ \bibnamefont {McCoy}},
  \bibinfo {author} {\bibfnamefont {J.~N.}\ \bibnamefont {Israelachvili}}, \
  and\ \bibinfo {author} {\bibfnamefont {I.}~\bibnamefont {Cohen}},\ }\href
  {\doibase 10.1126/science.1207032} {\bibfield  {journal} {\bibinfo  {journal}
  {Science}\ }\textbf {\bibinfo {volume} {333}},\ \bibinfo {pages} {1276}
  (\bibinfo {year} {2011})}\BibitemShut {NoStop}%
\bibitem [{\citenamefont {Habdas}\ \emph {et~al.}(2006)\citenamefont {Habdas},
  \citenamefont {Weeks},\ and\ \citenamefont {Lynn}}]{Habdas2006}%
  \BibitemOpen
  \bibfield  {author} {\bibinfo {author} {\bibfnamefont {P.}~\bibnamefont
  {Habdas}}, \bibinfo {author} {\bibfnamefont {E.~R.}\ \bibnamefont {Weeks}}, \
  and\ \bibinfo {author} {\bibfnamefont {D.~G.}\ \bibnamefont {Lynn}},\ }\href
  {\doibase 10.1119/1.2195396} {\bibfield  {journal} {\bibinfo  {journal} {The
  Physics Teacher}\ }\textbf {\bibinfo {volume} {44}},\ \bibinfo {pages} {276}
  (\bibinfo {year} {2006})}\BibitemShut {NoStop}%
\bibitem [{\citenamefont {Batchelor}(1967)}]{Bat67}%
  \BibitemOpen
  \bibfield  {author} {\bibinfo {author} {\bibfnamefont {G.}~\bibnamefont
  {Batchelor}},\ }\href@noop {} {\emph {\bibinfo {title} {An Introduction to
  Fluid Dynamics}}}\ (\bibinfo  {publisher} {Cambridge University Press},\
  \bibinfo {year} {1967})\BibitemShut {NoStop}%
  \bibitem [{\citenamefont {Ambari}\ \emph {et~al.}(1985)\citenamefont {Ambari},
  \citenamefont {Gauthier-Manuel},\ and\ \citenamefont {Guyon}}]{Ambari1985}%
  \BibitemOpen
  \bibfield  {author} {\bibinfo {author} {\bibfnamefont {A.}~\bibnamefont
  {Ambari}}, \bibinfo {author} {\bibfnamefont {B.}~\bibnamefont
  {Gauthier-Manuel}}, \ and\ \bibinfo {author} {\bibfnamefont {E.}~\bibnamefont
  {Guyon}},\ }\href {\doibase 10.1063/1.864990} {\bibfield  {journal} {\bibinfo
   {journal} {Physics of Fluids}\ }\textbf {\bibinfo {volume} {28}},\ \bibinfo
  {pages} {1559} (\bibinfo {year} {1985})}\BibitemShut {NoStop}%
\bibitem [{\citenamefont {Seckin}\ and\ \citenamefont
  {Kormaly}(1996)}]{Seckin1996}%
  \BibitemOpen
  \bibfield  {author} {\bibinfo {author} {\bibfnamefont {T.}~\bibnamefont
  {Seckin}}\ and\ \bibinfo {author} {\bibfnamefont {S.~M.}\ \bibnamefont
  {Kormaly}},\ }\href {\doibase 10.1021/ed073p193} {\bibfield  {journal}
  {\bibinfo  {journal} {Journal of Chemical Education}\ }\textbf {\bibinfo
  {volume} {73}},\ \bibinfo {pages} {193} (\bibinfo {year} {1996})}\BibitemShut
  {NoStop}%
\bibitem [{\citenamefont {Dolz}\ \emph {et~al.}(2005)\citenamefont {Dolz},
  \citenamefont {Delegido}, \citenamefont {Casanovas},\ and\ \citenamefont
  {Hernández}}]{Dolz2005}%
  \BibitemOpen
  \bibfield  {author} {\bibinfo {author} {\bibfnamefont {M.}~\bibnamefont
  {Dolz}}, \bibinfo {author} {\bibfnamefont {J.}~\bibnamefont {Delegido}},
  \bibinfo {author} {\bibfnamefont {A.}~\bibnamefont {Casanovas}}, \ and\
  \bibinfo {author} {\bibfnamefont {M.-J.}\ \bibnamefont {Hernández}},\ }\href
  {\doibase 10.1021/ed082p445} {\bibfield  {journal} {\bibinfo  {journal}
  {Journal of Chemical Education}\ }\textbf {\bibinfo {volume} {82}},\ \bibinfo
  {pages} {445} (\bibinfo {year} {2005})}\BibitemShut {NoStop}%
\bibitem [{\citenamefont {Courbin}\ \emph {et~al.}(2005)\citenamefont
  {Courbin}, \citenamefont {Cristobal}, \citenamefont {Winckert},\ and\
  \citenamefont {Panizza}}]{Courbin2005}%
  \BibitemOpen
  \bibfield  {author} {\bibinfo {author} {\bibfnamefont {L.}~\bibnamefont
  {Courbin}}, \bibinfo {author} {\bibfnamefont {G.}~\bibnamefont {Cristobal}},
  \bibinfo {author} {\bibfnamefont {M.}~\bibnamefont {Winckert}}, \ and\
  \bibinfo {author} {\bibfnamefont {P.}~\bibnamefont {Panizza}},\ }\href
  {\doibase DOI:10.1119/1.1949627} {\bibfield  {journal} {\bibinfo  {journal}
  {Am. J. Phys.}\ }\textbf {\bibinfo {volume} {73}},\ \bibinfo {pages} {851}
  (\bibinfo {year} {2005})}\BibitemShut {NoStop}%
\bibitem [{\citenamefont {Digilov}\ \emph {et~al.}(2008)\citenamefont
  {Digilov}, \citenamefont {Gurfinkel},\ and\ \citenamefont
  {Reiner}}]{Digilov2008}%
  \BibitemOpen
  \bibfield  {author} {\bibinfo {author} {\bibfnamefont {R.~M.}\ \bibnamefont
  {Digilov}}, \bibinfo {author} {\bibfnamefont {B.}~\bibnamefont {Gurfinkel}},
  \ and\ \bibinfo {author} {\bibfnamefont {M.}~\bibnamefont {Reiner}},\ }\href
  {\doibase DOI:10.1119/1.2904470} {\bibfield  {journal} {\bibinfo  {journal}
  {Am. J. Phys.}\ }\textbf {\bibinfo {volume} {76}},\ \bibinfo {pages} {968}
  (\bibinfo {year} {2008})}\BibitemShut {NoStop}%
\bibitem [{\citenamefont {Brizard}\ \emph {et~al.}(2005)\citenamefont
  {Brizard}, \citenamefont {Megharfi}, \citenamefont {Mah\'{e}},\ and\
  \citenamefont {Verdier}}]{Brizard2005}%
  \BibitemOpen
  \bibfield  {author} {\bibinfo {author} {\bibfnamefont {M.}~\bibnamefont
  {Brizard}}, \bibinfo {author} {\bibfnamefont {M.}~\bibnamefont {Megharfi}},
  \bibinfo {author} {\bibfnamefont {E.}~\bibnamefont {Mah\'{e}}}, \ and\
  \bibinfo {author} {\bibfnamefont {C.}~\bibnamefont {Verdier}},\ }\href
  {\doibase 10.1063/1.1851471} {\bibfield  {journal} {\bibinfo  {journal}
  {Review of Scientific Instruments}\ }\textbf {\bibinfo {volume} {76}},\
  \bibinfo {eid} {025109} (\bibinfo {year} {2005})}\BibitemShut {NoStop}%
\bibitem [{\citenamefont {Desbien}(2002)}]{Desbien2002}%
  \BibitemOpen
  \bibfield  {author} {\bibinfo {author} {\bibfnamefont {D.~M.}\ \bibnamefont
  {Desbien}},\ }\href@noop {} {\bibfield  {journal} {\bibinfo  {journal}
  {ProQuest Dissertations And Theses; Thesis (Ph.D.)--Arizona State
  University}\ } (\bibinfo {year} {2002})}\BibitemShut {NoStop}%
\bibitem [{\citenamefont {Lyman}(1981)}]{Lyman1981}%
  \BibitemOpen
  \bibfield  {author} {\bibinfo {author} {\bibfnamefont {F.}~\bibnamefont
  {Lyman}},\ }in\ \href@noop {} {\emph {\bibinfo {booktitle} {Mainstreaming
  Digest: A Collection of Faculty and Student Papers}}},\ \bibinfo {editor}
  {edited by\ \bibinfo {editor} {\bibfnamefont {A.~S.}\ \bibnamefont
  {Anderson}}}\ (\bibinfo  {publisher} {The University of Maryland},\ \bibinfo
  {year} {1981})\BibitemShut {NoStop}%
\bibitem [{\citenamefont {Guerra}\ \emph {et~al.}(2011)\citenamefont {Guerra},
  \citenamefont {Corley}, \citenamefont {Giacometti}, \citenamefont {Holland},
  \citenamefont {Humphreys},\ and\ \citenamefont {Nicotera}}]{Guerra2011}%
  \BibitemOpen
  \bibfield  {author} {\bibinfo {author} {\bibfnamefont {D.}~\bibnamefont
  {Guerra}}, \bibinfo {author} {\bibfnamefont {K.}~\bibnamefont {Corley}},
  \bibinfo {author} {\bibfnamefont {P.}~\bibnamefont {Giacometti}}, \bibinfo
  {author} {\bibfnamefont {E.}~\bibnamefont {Holland}}, \bibinfo {author}
  {\bibfnamefont {M.}~\bibnamefont {Humphreys}}, \ and\ \bibinfo {author}
  {\bibfnamefont {M.}~\bibnamefont {Nicotera}},\ }\href {\doibase
  DOI:10.1119/1.3555507} {\bibfield  {journal} {\bibinfo  {journal} {The
  Physics Teacher}\ }\textbf {\bibinfo {volume} {49}},\ \bibinfo {pages} {175}
  (\bibinfo {year} {2011})}\BibitemShut {NoStop}%
\end{thebibliography}
\end{document}